\documentclass[5p,times]{elsarticle}
\usepackage{amsmath}
\usepackage{amsfonts}
\usepackage{epsfig}
\usepackage{graphics} 
\usepackage[dvipsnames]{xcolor}
\usepackage{placeins}
\usepackage[algo2e,linesnumbered,commentsnumbered, lined, boxed,ruled,vlined]{algorithm2e}
\definecolor{cppColorBackground}{rgb}{1.,1.,1.}
\definecolor{cppColorComment}{rgb}{0.0,0.67,1.}
\definecolor{cppColorLine}{rgb}{0.6,0.6,0.6}
\definecolor{cppColorString}{rgb}{0,0.501,145}
\definecolor{cppColorKey}{rgb}{0.5,0.5,0}
\definecolor{cppColorDigit}{rgb}{0,0,0.5}
 \usepackage{lmodern}
 \usepackage{listings}
\begin{document}
\begin{frontmatter}
\title{Optimizations of the Eigensolvers in the ELPA Library}
\author[MPCDF]{P. K\r{u}s}
\ead{pavel.kus@mpcdf.mpg.de}
\author[MPCDF]{A. Marek}
\author[TUM]{S. S. K\"ocher}
\author[FHI]{H.-H. Kowalski}
\author[FHI]{C. Carbogno}
\author[TUM]{Ch. Scheurer}
\author[TUM]{K. Reuter}
\author[FHI]{M. Scheffler}
\author[MPCDF]{H. Lederer}
\address[MPCDF]{Max Planck Computing and Data Facility, D-85747 Garching, Germany}
\address[TUM]{Theoretical Chemistry and Catalysis Research Center, Technische Universität München, D-85747 Garching, Germany}
\address[FHI]{Fritz-Haber-Institut der Max-Planck-Gesellschaft, D-14195 Berlin, Germany}

\begin{abstract}
	The solution of (generalized) eigenvalue problems for symmetric or Hermitian matrices is a common subtask of many 
	numerical calculations in electronic structure theory or materials science. Depending on the scientific problem,
	solving the eigenvalue problem can easily amount to a sizeable fraction of the whole numerical calculation, and quite often is even the
	dominant part by far. For researchers in the field of computational materials science, an efficient and scalable solution of the 
	eigenvalue problem is thus of major importance.
	The ELPA-library (\emph{E}igenvalue \emph{S}o\emph{L}vers for \emph{P}etaflop-\emph{A}pplications) is a well-established dense direct 
	eigenvalue solver library, which has proven to be very efficient and scalable up to very large core counts. It is in a wide-spread production use on
	a large variety of HPC systems worldwide, and is applied  by many codes in the field of materials science.
	In this paper, we describe the latest optimizations of the ELPA-library for new HPC architectures of the Intel Skylake processor family
	with an AVX-512 SIMD instruction set, or for HPC systems accelerated with recent GPUs.
 	Apart from those direct hardware-related optimizations, we also describe a complete redesign of the API in a modern modular
	way, which, apart from a much simpler and more flexible usability, leads to a new path to access system-specific performance optimizations.
	In order to ensure optimal performance for a particular scientific setting or a specific HPC system, the new API allows the user to influence 
	in straightforward way the internal details of the algorithms and of performance-critical parameters used in the ELPA-library. 
	On top of that, we introduced an autotuning functionality, which allows for finding the best settings in a self-contained automated way, without
	the need of much user effort. 
	In situations where many eigenvalue problems with similar settings have to be solved consecutively, the autotuning process of the ELPA-library can be done ``on-the-fly'', without the need of preceding the simulation with an ``artificial'' autotuning step. 
	Practical applications from materials science which rely on reaching a numerical convergence limit by so-called self-consistency iterations can profit from the on-the-fly autotuning.
	On some examples of scientific interest, simulated with the FHI-aims~\cite{Blum:2009fe} application, the advantages of the latest optimizations of the ELPA-library are demonstrated.
\end{abstract}

\end{frontmatter}
\section{Introduction}

When developing and maintaining a library for HPC applications the developers generally have to make a difficult decision: on the one hand they
can decide to develop a \emph{specialized} library which shows best performance on a certain HPC hardware, on the other, they can choose to develop a \emph{general} library which 
supports a huge variety of HPC systems, albeit, as a consequence, the performance tuning becomes much more complex. The reason for this choice is mandated by the extreme 
variety of available HPC systems: it is an almost impossible endeavour to optimize a library for each available processor from different manufacturers (or even different 
processors from the same manufacturer) each with its own characteristics of CPU frequency, cache hierarchy and cache sizes, SIMD instructions set, only to mention a few. This 
task is further complicated by the availability of different interconnects, memory configurations and, additionally, the possible presence of GPU (or other) accelerators.
In the HPC community this led to the fact that there are on the one hand \emph{general}, \emph{standard} libraries, like the famous BLAS, LAPACK, and ScaLAPACK~\cite{blas, lapack, scalapack}, which are 
open-source and can be  compiled and used on every system.  And on the other hand, there exist,  normally as closed source, vendor-provided \emph{optimized implementations} of these 
standard libraries, which run with very good performance on the vendor-supplied systems. Some typical examples for vendor-specific implementations of BLAS, LAPACK, and ScaLAPACK are Intel's MKL~\cite{mkl}, IBM's (p)essl~\cite{essl, pessl} and Nvidia's cublas~\cite{cublas} libraries.

In this paper we present our approach to provide optimal performance on different HPC systems.
Originally started in 2008, the ELPA library is nowadays one of the most used HPC libraries for solving symmetric (or hermitian) eigenvalue problems. 
It is installed
at many HPC centers in the world and used by many important applications in the field of material structure theory and molecular dynamics, like Gromacs~\cite{gromacs}, FHI-aims~\cite{Blum:2009fe}, Quantum-Esspresso~\cite{quantumesspresso}, and Wien2k~\cite{wien2k}, just to mention a few.

Over the years, during the development of the ELPA library, a lot of attention has been paid to search for optimal algorithms, node-level optimization
(efficient usage of BLAS level 3 routines, optimal blocking for cache, the development of optimized  kernels using compiler intrinsics or even
assembly) and very efficient MPI communication patterns. Some of the results were already published in previous publications
\cite{elpa_2011} and~\cite{elpa_2014}. 


Apart from these specialized, traditional-style optimizations, which is most likely very similar to the approaches of a hardware vendor implementing optimized versions of a library, in ELPA 
we additionally add another class of optimizations, which aims to help the performance portability on different processors and architectures, ranging from a local PC to possibly all HPC systems.
It is reasonable to assume that it will not be possible to find one implementation which is the best-performing in all conceivable
scenarios. Thus, in order to deliver excellent performance for each scenario, our approach is based on having several variants of an algorithm, 
or on allowing to fine-tune the behavior (e.g. tuning different block sizes in different parts of one algorithm) of a particular algorithm and to allow a \emph{run-time choice} of the 
best-performing settings. The idea is to provide reasonable default settings, which work excellent, or at least reasonably well in most cases, but our approach also offers two additional
possibilities to tweak the execution of the library: one the one hand, users with a deep domain knowledge are allowed to change some (or even all) run-time choices in a very clear 
and easy way. On the other hand we have implemented an autotuning functionality, which allows less experienced users to find the best settings for their specific combination of problem size 
and used hardware.

As a consequence, a complete redesign of the ELPA API has been done, which allows the user (if needed) to fully control all possible settings in order to influence the time-to-solution of the
ELPA library. Furthermore, the new API allows in a natural way to implement a (semi) automated search for the best settings. Last but not least, the new API allows to introduce new tunable 
parameters in the future without changing the API and breaking compatibility to previous version of the ELPA library.

This paper is organized as follows: After recapitulating the mathematical background of the solution to a (generalized) eigenvalue problem in Section~\ref{sec:EVP}, we describe in 
Section~\ref{sec:architecture_optimization} the latest optimization for the Intel Xeon Skylake and NVIDIA GPU architectures and show some performance results. In Section~\ref{sec:redesign} 
we introduce the new API and the autotuning capability of the ELPA library and explain the advantages and the usage of the new 
approach. Finally, using two real-world examples
in Section~\ref{sec:applications}, we show  the benefits the applications can get by using 
the latest version of the ELPA library. The paper is then concluded 
in Section~\ref{sec:conclusions}.

\section{The eigenvalue problem solved by ELPA}
\label{sec:EVP}
We look for the solution of a (possibly generalized) eigenvalue problem (EVP)
\begin{equation}
 AV=BV\Lambda.
\end{equation}
The steps of finding a solution to this problem are well known and conceptually simple, see, e.g, \cite{golub}.
The basic ELPA algorithm has already been described in \cite{elpa_2011}, \cite{elpa_2014} or \cite{elpa_gpu_2017}, we will, 
however, recall the basic steps for the sake of completeness.
First, if we are to compute a generalized EVP, we start by computing the Cholesky decomposition 
\begin{equation}
\label{eq:cholesky}
 B=LL^H
\end{equation}
 of a possibly complex matrix $B$ ($L^H$ denoting the conjugate transpose of $L$) 
 and by transforming the problem to a standard one,  $\tilde{A}=\tilde{V}\Lambda$  with  
\begin{equation}
\label{eq:trans_gen_forw}
\tilde{A} = L^{-1}A(L^{-1})^{H}.
\end{equation}
More information regarding the new API for the generalized EVP can be found in Section~\ref{sec:generalized}.
In the case that a standard EVP is to be solved, the previous step is skipped.
In any case, the next step is the reduction of the matrix to a tridiagonal form
\begin{equation}
T=Q\tilde{A}Q^{H},
\label{eq:tridiagonalization}
\end{equation}
where $Q=Q_n\cdots Q_2 Q_1$ and $Q^T=Q_1^{H}Q_2^{H}\cdots Q_n^{H}$
are the successive
Householder matrices reducing one column of $\tilde{A}$ at a time. The Householder matrices 
$
 Q_i = I - \beta_i v_i v_i^{H}
$
are never constructed explicitly, but are always represented only by the Householder vector $v_i$.
In each step, a new Householder vector is computed and stored in place of an eliminated column of $A$, reducing the memory
requirements. The diagonal and sub-diagonal of the resulting matrix are stored separately.
Applying transformations on $A$ from both sides complicates blocking and usage of efficient BLAS level 3 kernels.
This restriction is alleviated in the two-stage algorithm, which will be briefly described later. 
The next step is the solution of the tridiagonal eigenvalue  problem 
\begin{equation}
 T\hat{V} = \hat{V}\Lambda
 \label{eq:Esolv}
\end{equation}
and the final step is the back transformation of the $k$ required eigenvectors, the last step (\ref{eq:trans_gen_back})
being performed only 
for the generalized EVP:
\begin{eqnarray}
 \tilde{V}&=&Q^H\hat{V} \\
 \label{eq:trans_gen_back}
  {V}&=&(L^{-1})^{H}\tilde{V}.
\end{eqnarray}


The two stage algorithm, featured in ELPA 2, differs from the previous description in performing 
the conversion to the tridiagonal matrix in two steps. In the first step, the matrix is reduced to a banded form. 
This allows usage of highly optimized BLAS level 3 functions. In the second step, the matrix is further reduced to 
the tridiagonal form. The two stage tridiagonalization is almost always faster, however, the price to pay is 
the need to also transform each eigenvector (found by the tridiagonal solver) twice, in order to find the eigenvector
of the original system. This makes ELPA 2 an obvious choice when only a small part of the eigenvectors are sought. When most or all of the eigenvectors are needed, the best choice might depend on other parameters
(such as the matrix size, particular hardware, etc) as well.

\begin{figure}
\begin{center}
\includegraphics[width=0.5\textwidth]{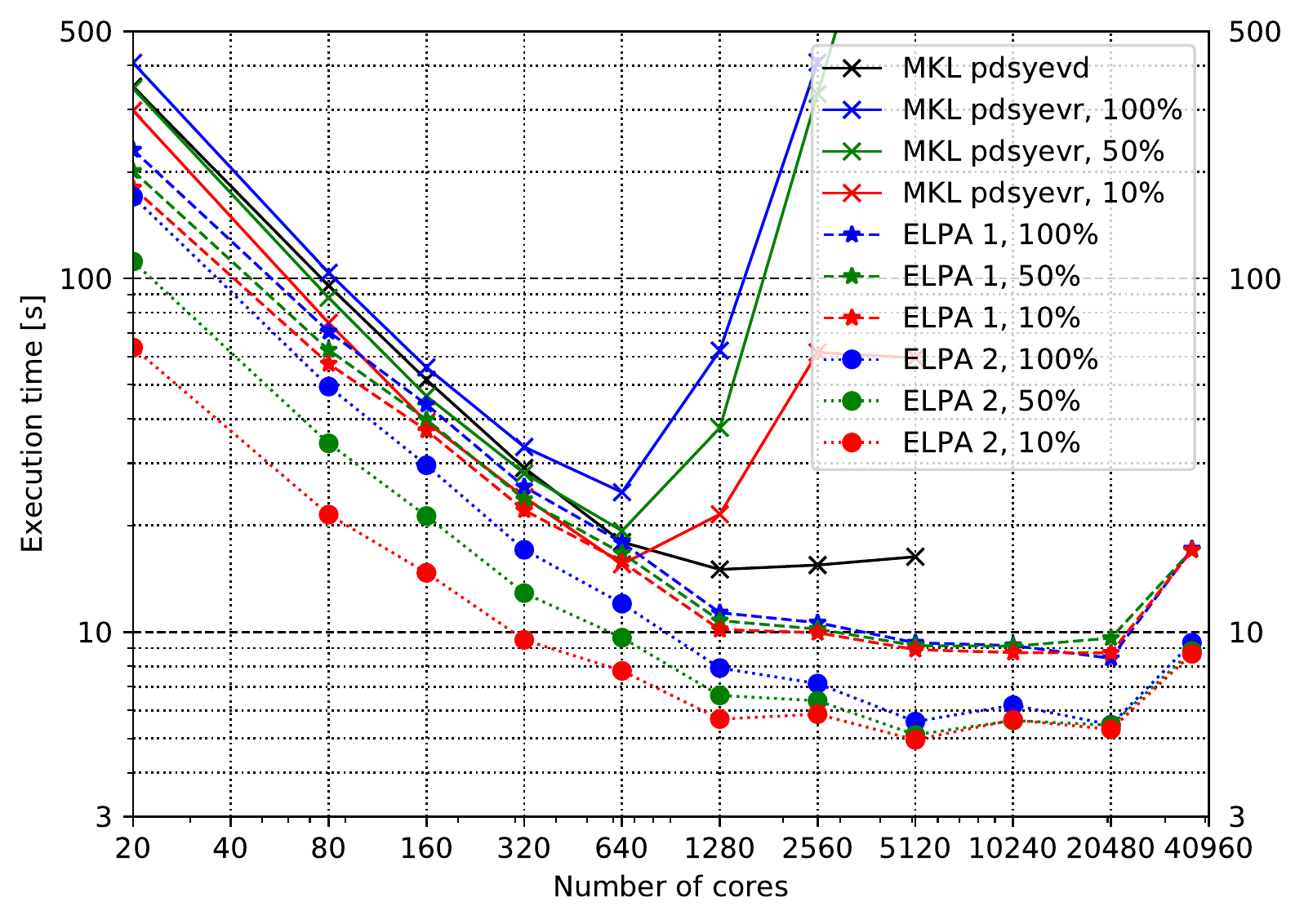}
	\caption{\label{fig:hydra} Scaling of the ELPA library on the ''Hydra'' system at MPCDF, an Intel Xeon Ivy Bridge system. For comparison results for Intel's MKL 2017 are shown, using the MKL implementation of the routine ''pdsyevd'' (all eigenvectors are computed), and the routine ''pdsyevr'' (part of the eigenvectors are computed). All the computations use a double-precision, real matrix of size $n=20000$.}
\end{center}
\end{figure}

\section{Optimizations for modern architectures}
\label{sec:architecture_optimization}

\subsection{Intel Xeon Skylake optimizations}
\label{ssec:skylake}
%
%
We present some scaling results of the ELPA library obtained
using the two last generations of the large HPC systems at MPCDF. 
As of the time of writing, the previous generation supercomputer ``Hydra'' (3500 two-socket Ivy Bridge nodes 
with CPU frequency 2.8 GHz and 20 cores each connected via an InfiniBand interconnect) is being replaced by the new Skylake-based system ``Cobra'' (currently 3188 two-socket Skylake nodes with CPU frequency 
2.4 GHz and 40 cores each connected via an OmniPath network). 

\begin{figure}
\begin{center}
\includegraphics[width=0.5\textwidth]{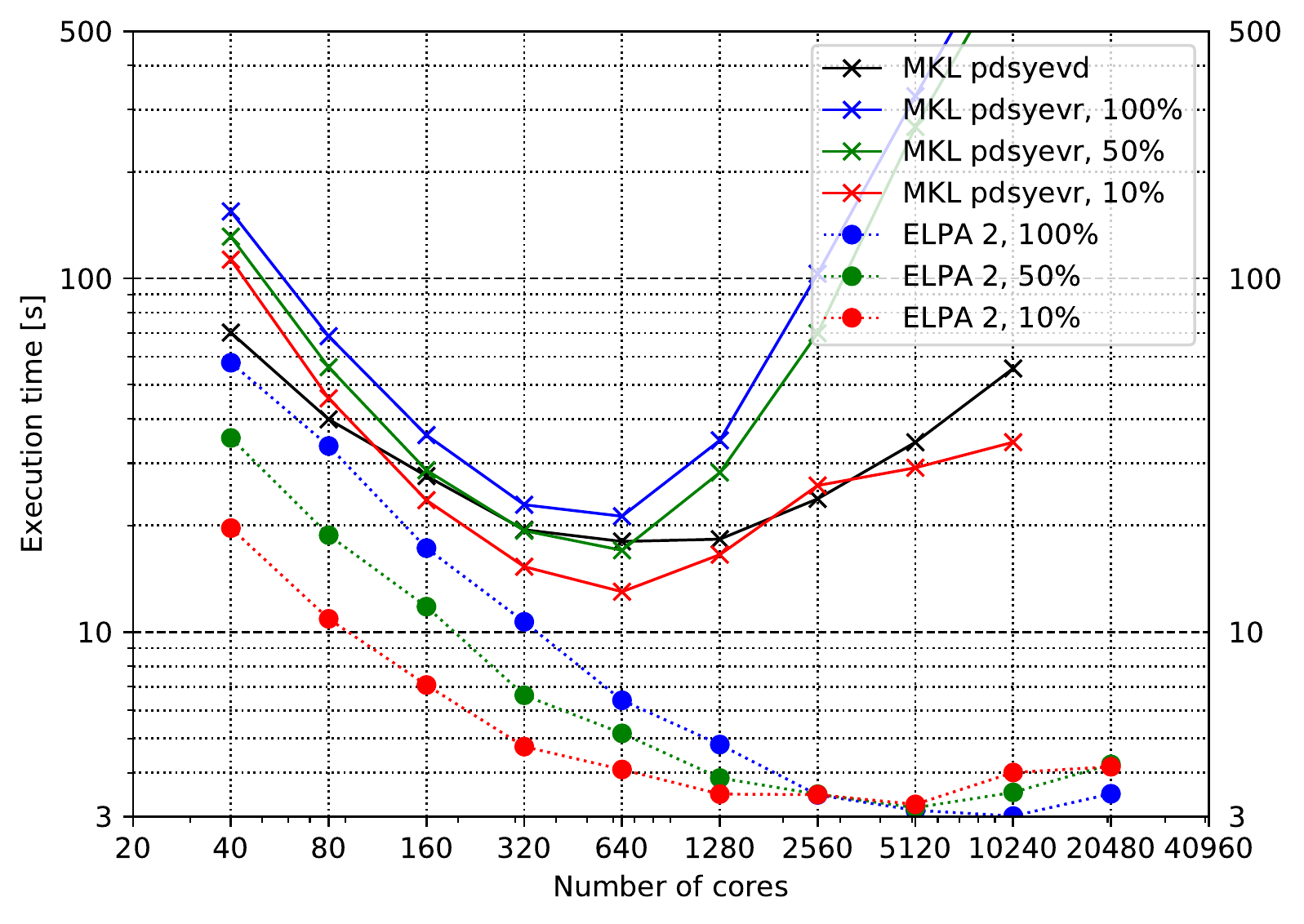}
 \caption{\label{fig:cobra} Scaling of the ELPA library on the ``Cobra'' HPC system at MPCDF, an Intel Xeon Skylake system. A comparison with Intel's MKL 2018 version is shown. The setting are identical to the one in Figure~\ref{fig:hydra}. All test programs have been compiled with the Intel 2018 compiler and use Intel's 2018 MPI library and MKL.}
\end{center}
\end{figure}

Even though both the ELPA and the MKL library can handle complex matrices, or single-precision numbers, for sake of simplicity, all the comparisons
presented in this section were done with double-precision real numbers. Nonetheless, similar results are expected in single-precision and/or complex calculations.
In Fig.~\ref{fig:hydra} we show results from the Ivy Bridge machine Hydra to serve as a baseline.
%
While for a smaller number of cores ($<$640) the MKL routines are competitive with the 1-stage ELPA solver, the ELPA 2-stage solver always shows considerably better performance.
Furthermore, the MKL implementation shows a worse scaling behaviour beyond 1000 cores and the performance deteriorates significantly, whereas this is not the case for both the ELPA 1 and ELPA 2 solvers.
Thus an application using ELPA at large core counts will suffer much less from the slight loss of scalability.
We want to point out that at the core count where also for ELPA 1 and ELPA 2 the performance worsens, the local matrix has a size of 100x100 only -- very small indeed, which might lead to a too large communication overhead and thus the breakdown of the scaling.

\begin{figure}
\begin{center}
\includegraphics[width=0.5\textwidth]{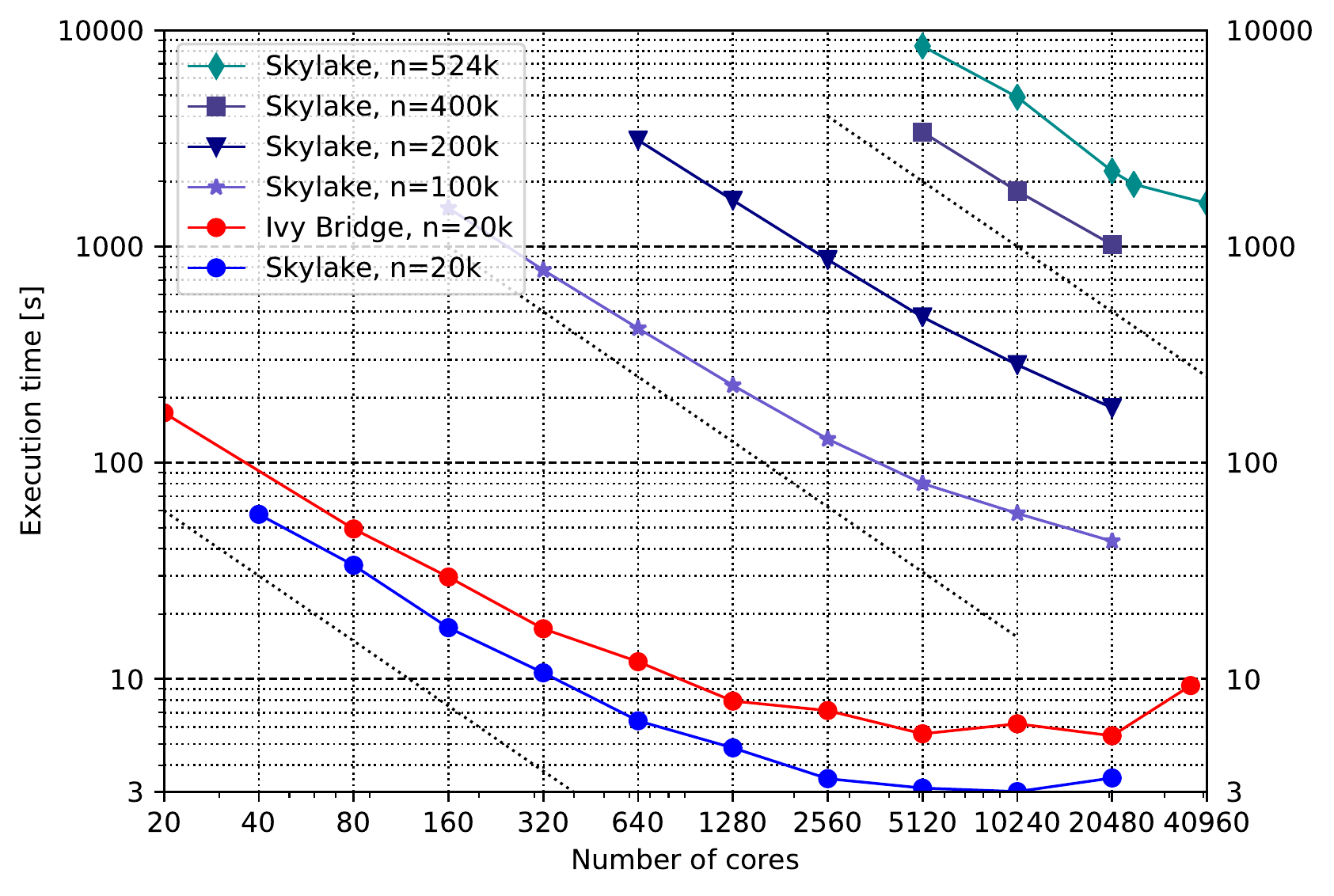}
 \caption{\label{fig:cobra_hydra} Scaling of the ELPA library for real, double precision matrices and 100\% eigenvectors sought. We 
 show a direct comparison of the performance between an Ivy Bridge and a Skylake based HPC system for matrix size
 $n=20000$ (both curves have already been seen in Figures~\ref{fig:hydra} and \ref{fig:cobra}). 
 Additionally, scaling curves  for larger matrices on the Skylake system are also shown together with dotted lines indicating
 an ideal scaling.}
\end{center}
\end{figure}

In Fig.~\ref{fig:cobra} we show the latest results from the Skylake machine Cobra.
One can see that both libraries perform considerably better than on the previous machine (see Fig.~\ref{fig:hydra}), 
but the general results of the scalability and the comparison between ELPA and the MKL library are the same.
The scaling behaviour of ELPA for matrices of different size (up to over half million) on the Cobra machine is shown in Fig.~\ref{fig:cobra_hydra}. Very good scaling is demonstrated throughout the investigated region.
The same figure also shows a  direct comparison between the performance of ELPA 2 on Ivy Bridge and Skylake for matrix size 20000.
Most of the performance benefits of the newer Intel hardware are automatically delivered by the use of a 
recent compiler supporting the latest Xeon processors and an optimized implementation of BLAS in Intel's MKL library. 
In the ELPA 2 solver, however, there is a specific part of the back transformation algorithm, which is very computationally intensive, 
but cannot be efficiently implemented by calls to the BLAS library. In order to achieve optimal performance, these small but important computational kernels have been implemented with compiler SIMD intrinsics, or even with assembly language. Such implementations are not portable and thus it is necessary to write a new variant
for each new instruction set, as it has already been described in \cite{elpa_2014}.

Due to its 512 bit wide vector registers, the latest SIMD instruction set of Intel Xeon processors, known as AVX-512, can 
process four double or eight single precision real values (2 and 4, respectively, in case of complex numbers) 
in one vector instruction. Theoretically, this gives twice the performance than the older 256 bit AVX-2 
vector instructions. However, due to heat management, 
modern Xeon processors utilize CPU frequency throttling, which occurs when the computations are too intensive and the processor 
starts to heat up. For the user it is far from trivial to understand, when this thermal frequency throttling 
becomes active and at which ``AVX frequency'' the vector instructions are executed. The theoretical factor of two 
performance gain of the hand-tuned AVX-512 kernels can thus be reduced. 

The overall effect can be 
judged from Figure~\ref{fig:cobra_hydra}, which shows results using the AVX-2 (on Ivy Bridge) and AVX-512 (on Skylake)
hand written kernels, 
respectively. Note, however, that on both machines the ``effective'' frequency at which the vector instructions 
were executed are not necessarily the same and a speedup of two is usually not achieved. 
Results show a speedup in the range of roughly 1.5x (80 cores) and 2x (10240 cores),
%
which is a very good achievement, given that a large part
of the performance improvement of the new HPC systems is delivered by an increasing number of cores, which is not accounted for
in this comparison. Indeed, if we compared the performance based on the number of nodes instead of the number of cores, the improvement for 
Skylake nodes (40 cores/node) over Ivy bridge (20 cores/node), would be even higher. 

\subsection{GPU-related optimizations}
\label{ssec:gpu}

\begin{figure}
\begin{center}
 \includegraphics[width=0.5\textwidth]{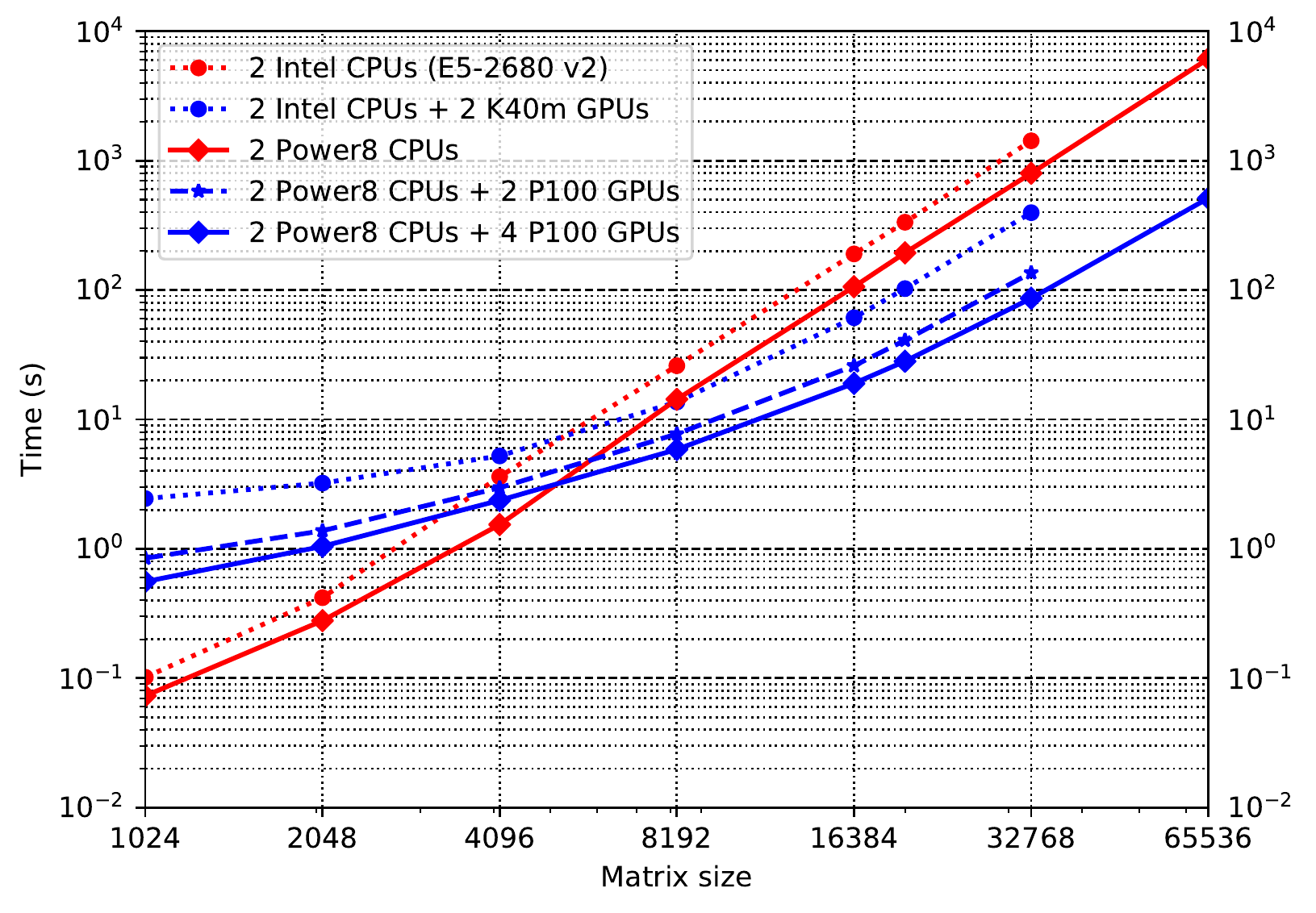}
 \caption{\label{fig:cpu_gpu}Total solution times in seconds for CPU only and CPU + GPU versions of ELPA 1. Results from two 
 different architectures are shown for comparison. The problem with the largest matrix is not always computed due to 
 memory limitations of the GPU cards.}
\end{center}
\end{figure}

Modern GPU cards offer a very large peak performance, which  exceeds the peak performance of the whole CPU node
significantly. It is thus important to be able to offload arithmetically most intensive parts of the computation to the 
GPU card.
Since the CPU version of the ELPA library uses BLAS calls for local matrix and vector calculations, 
the incorporation of the  GPU utilisation is quite straightforward, as it has already been described in detail in~\cite{elpa_gpu_2017}.
%
When appropriate, blocks of locally owned matrices and vectors are explicitly transferred between the host memory
and the device memory and the  calculations are then performed on the GPU devices  
using calls to a highly optimized cuBLAS library\cite{cublas}. 
The GPU implementation thus still relies on the very efficient MPI-based implementation of ELPA with 
a block-cyclic distribution of the matrices, but, on top of that,  each MPI task communicates
with the GPU in order to speed-up arithmetically intensive local computations.

Despite our effort to keep a similar code path for both the CPU and the GPU version, 
occasionally  more substantial changes in the algorithm had to be done in order to obtain the best performance. 
The CPU version of ELPA has been highly optimized by keeping the cache reuse in mind. For this reason, many of the algorithms 
use explicit blocking and try to reuse pieces of the matrices, which are in cache, for multiple operations. 
This is often not favorable for a GPU, since it cannot benefit from caching, but, rather, it benefits from 
large amounts of data being processed in one run. Some of the blocking strategies thus had to be changed and 
the algorithm had to be altered to better suite the GPU.
%
%
%
In a typical compute node, the number of CPU cores (several dozen) is much bigger then the number of GPU devices
and thus in order to ensure good performance, the Nvidia Multi-process Service \cite{MPS} 
has to be  used to dispatch the requests from individual MPI tasks to the GPU devices and to use its streams efficiently.

We have tested the GPU implementation on two different systems representing two different architectures:
the first one is a rather old, but standard Intel-based 
system with two Intel Xeon Ivy Bridge processors  with 20 cores in total,
equipped with two Nvidia Tesla K40m GPUs, connected with PCIExpress Gen2 to the host.
The second machine has  two IBM Power8 processors 
and four  powerful Nvidia Pascal P100 GPUs and NVLINK interconnect. 
On the Intel system we used the Intel 16 compilers, MKL 2017 (containing the BLAS functions) and CUDA 8 (containing the cuBLAS 
implementation). On the Power8 machine we used the GNU compilers version 5.3, the ESSL library version 5.5 (containing the BLAS 
functions) and CUDA 9.

%
%
\begin{figure}
\begin{center}
 \includegraphics[width=0.5\textwidth]{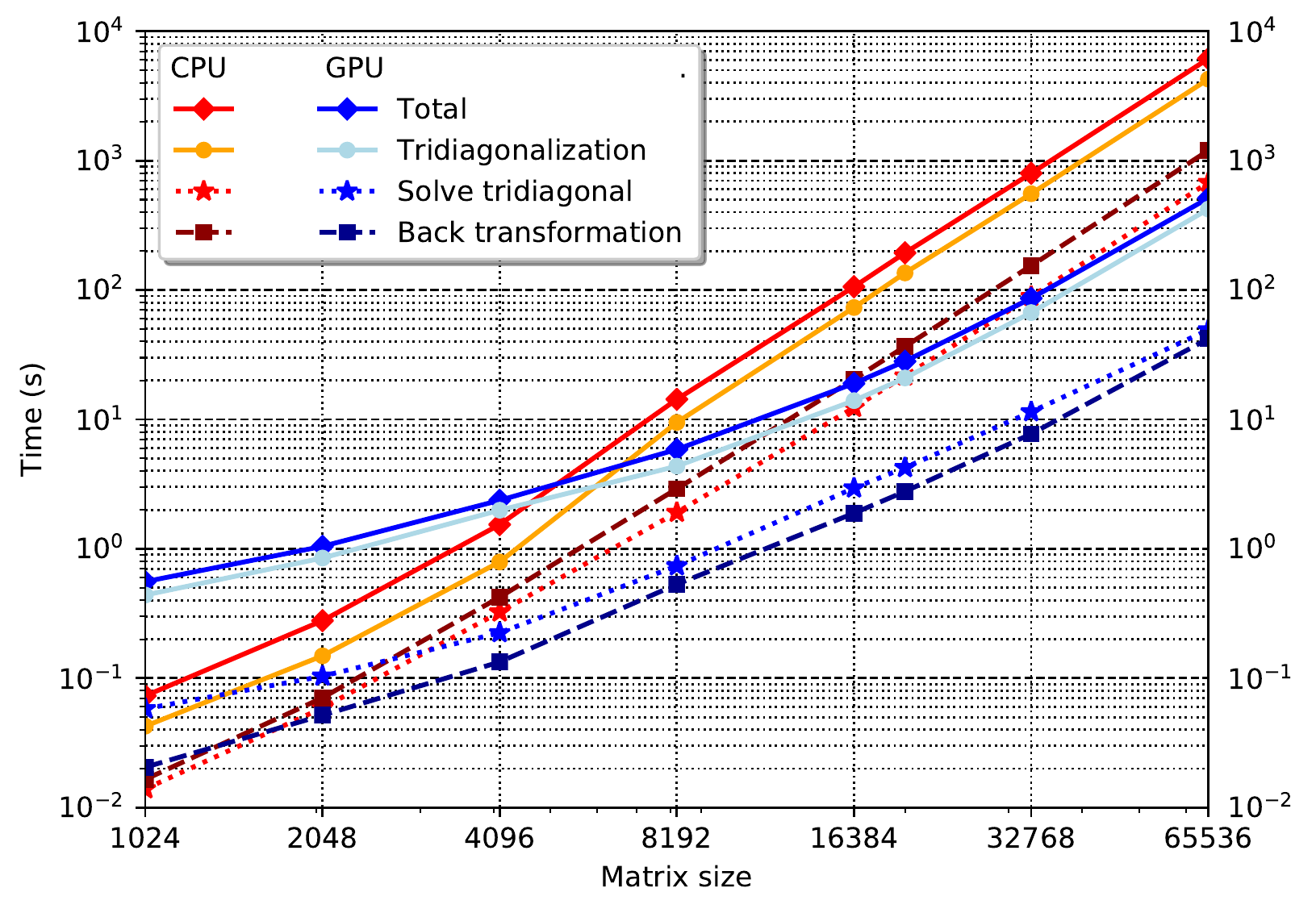}
 \caption{\label{fig:cpu_gpu_detailed}Detailed comparison of individual parts of the ELPA 1 algorithm on the Power8 node (see in text)
 using CPU only and CPU and all 4 GPUs.} 
\end{center}
\end{figure}

Figure~\ref{fig:cpu_gpu} shows the comparison of the total run-time for problems with different matrix sizes  
on both mentioned machines and for both the CPU and the GPU versions of ELPA 1.
On the  Power8 machine, the CPU version is generally faster, mostly due to higher
frequency (4 GHz compared to 2.8 GHz of the Intel machine). 
For both machines and small matrices,
the CPU implementation is significantly faster, since 
there is not enough workload to saturate the GPUs.
From a certain threshold on (matrix size of around 5000),
however, this behavior changes and the GPU version becomes much faster. 
The maximal speed-up for the largest computed matrices is 3.6x on the Intel + K40m system and 11.9x on the Power8 + P100 
system, respectively.
%
For multiple-node runs, the performance benefit of the GPU version is comparable if the local matrix size per node 
is of a similar size as in the described setup.

Going into more details, we show 
in Figure~\ref{fig:cpu_gpu_detailed} the breakdown of the runtime into  individual steps
on the Power8 node (CPU only, and CPU with 4 GPUs).
In both cases, the compute-time is dominated by the tridiagonalization step. This step contains 
BLAS level 2 operations (matrix-vector multiplications), which can not be very efficiently implemented in either BLAS (CPU)
or cuBLAS (GPU) libraries. Still, since most of the work done in both the tridiagonal solver and the back 
substitution is hidden in BLAS level 3
operations, which are particularly efficient on GPU, we can see that 
the GPU version is even more limited by the BLAS level 2 dominated 
tridiagonalization.

The solution times for large matrices are listed in detail in Table~\ref{tab:cpu_gpu_detailed}. The first two cases are intended for 
a comparison between the two tested architectures, since the same matrix size is used, and only 2 GPUs are utilized for the Power8 
system. The last case shows the largest possible (due to memory limitations)
computed matrix with the full Power8 node (using all 4 GPUs). It is
worth noticing, that for the back substitution step we get almost 30x speed-up. This is caused by the efficient implementation 
using BLAS level 3 operations only. Indeed, even the total speed-up of almost 12x is very good. 

\begin{table*}
\begin{center}
\caption{\label{tab:cpu_gpu_detailed}Performance of individual stages of the ELPA 1 algorithm.}
\begin{tabular}{c|c|cc|c|cc|c|cc}
\hline
machine & \multicolumn{3}{|c|}{2xIntel+2xK40m}  & \multicolumn{3}{|c|}{2xPower8+\textbf{2x}P100 } & \multicolumn{3}{|c}{2xPower8+\textbf{4x}P100}  \\ 
mat. size  & \multicolumn{3}{|c|}{32768} & \multicolumn{3}{|c|}{32768}  & \multicolumn{3}{|c}{65536}  \\
\hline
{}  &CPU & \multicolumn{2}{c|}{GPU} & CPU & \multicolumn{2}{c|}{GPU}  & CPU & \multicolumn{2}{c}{GPU}  \\
{} & t(s) & t(s) & s-up & t(s) & t(s) & s-up & t(s) & t(s) & s-up   \\
\hline
Total & 1424  & 396  & \textbf{3.6x} & 798 & 136 & \textbf{5.9x}& 6139& 514& \textbf{11.9x} \\ 
Tridiag. & 1108 & 333 & \textbf{3.3x} & 555 & 113 & \textbf{4.9x} & 4263& 422& \textbf{10.1x} \\
Solve & 117 & 24.1 & \textbf{4.9x} & 88.9 & 12.9 & \textbf{6.9x} & 678& 49.2& \textbf{13.8x} \\ 
Back s. & 198 & 36.9 & \textbf{5.4x} & 154 & 10.3  & \textbf{15x}  & 1198& 42.1& \textbf{28.5x} \\ 
\hline
\end{tabular}
\end{center}
\end{table*}

\section{Redesign of the ELPA library}
\label{sec:redesign}
The ELPA library has been developed for some time already. It's original interface had been inspired by the interfaces of the famous HPC libraries BLAS,
LAPACK, and SCALAPACK. This means there was one specific function call for each purpose, and the user had to specify all the input and output data, 
but also all parameters controlling the behaviour in the function signature. While this is reasonable for the above mentioned standard libraries,
where the signatures of each function are essentially fixed since a long time, this turned out not to be optimal for the ELPA library:
firstly, adding new functionality to the ELPA library always implied changing the function signatures and thus breaking the API and ABI compatibility
between different releases, which is not acceptable from the users point of view. Secondly, over time the signatures of the functions become just
too long and thus error prone.

It was thus decided to redefine the API of the ELPA library in a new way such that, firstly, the function signatures are as simple as possible. 
Secondly, the new API design introduces flexibility, such that adding a new parameter to the library does not change the signatures (see Subsection~\ref{ssec:redesign_object}). By this simplification of the API, it has been possible to 
expose many internal parameters of the library, which allow for the run-time performance tuning by the calling programs. Furthermore, as discussed
further below, this allows for an autotuning of these parameters in an easy way (to be discussed in Subsection~\ref{sec:autotuning}), which would have been impossible with the old rigid interface.
\begin{figure}
\begin{lstlisting}[label={code:main}, caption={Example use of the ELPA object}]
use elpa
class(elpa_t), pointer :: e
integer :: success
e => elpa_allocate()
! set parameters describing the matrix and 
! its MPI distribution, they are required
call e%set("na", na, success)
call e%set("nev", nev, success)
call e%set("local_nrows", na_rows, success)
call e%set("local_ncols", na_cols, success)
call e%set("nblk", nblk, success)
call e%set("mpi_comm_parent", mpi_comm_world, success)
call e%set("process_row", my_prow, success)
call e%set("process_col", my_pcol, success)
success = e%setup()
! if desired, set other run-time options
call e%set("solver", elpa_solver_2stage, success)
! call one of the solution methods
! the data types of a, ev, and z determine whether
! it is single/double precision and real/complex
call e%eigenvectors(a, ev, z, success)
! cleanup
call elpa_deallocate(e)
call elpa_uninit()
\end{lstlisting}
\end{figure}




\subsection{Examples of the new API}
\label{ssec:redesign_object}
We have chosen the object-oriented approach (using the modern Fortran language)
to achieve all the previously mentioned goals. In the simplest variant, 
the user can simply create the ELPA object, set the required problem properties, call the solution method and then free the object again.
A sketch of such code is shown in Code~\ref{code:main} for a Fortran and in Code~\ref{code:main_C} for an C implementation.
When re-designing the ELPA API, great care has been taken that for the user the transition from the old to the new API is easily done 
without a lot of effort. We want to stress that no changes to the data structures of the input and output data to ELPA library have been
done and the users do not have to change their applications dramatically.
As already mentioned, in the new API adding new parameters is quite flexible.
Since all parameters are identified by strings and handled by the \emph{set} and \emph{get} methods, adding a new parameter does not require any change of API.
By defining for each parameter an implicit value, backward compatibility between different versions of the library is ensured.
The library comes with methods, which provide functionality to print, save and load the status of an ELPA object (and all it's parameters).
%

\begin{figure}
\begin{lstlisting}[language=C, label={code:main_C}, caption={Example use of the C interface}]
#include <elpa/elpa.h>
elpa_t handle;
handle = elpa_allocate(&error);
// set everything like in the Fortran example,
// using the handle
elpa_set(handle, "na", na, &error);
// values can also be retrieved
elpa_get(handle, "solver", &value, &error);
printf("Solver is set to %d \n", value);
// solve the EV problem
// the data types of a, ev, and z determine whether 
// it is a single/double, real/complex problem
elpa_eigenvectors(handle, a, ev, z, &error);
// cleanup
elpa_deallocate(handle);
elpa_uninit();
\end{lstlisting}
\end{figure}



The new API also allows for having multiple instances of ELPA at the same time, each of them possibly in a different state or tuned for different type of problems. 
Of course, all instances can be created and destroyed at will independently of each other.
For example, it is thus possible to have two instances, where the first one is configured to do all the computations on the CPUs and the second one is
configured to use GPUs. This might be beneficial when a user application has two classes of 
eigenvalue problems to solve: one for small and one for large matrices. Another example might be having one ELPA instance
optimized for single precision and an other one for double precision calculations if the application wants to mix the two. Such 
example, and the performance gained with this approach is described in Section~\ref{ssec:mixed_precision}. 
%

\begin{figure}
\begin{lstlisting}[label={code:generalized}, caption={Generalized EVP API}]
! for the generalized EVP a boolean parameter 
! is_already_decomposed for the first time with 
! the same matrix b, call with .false.
call e%generalized_eigenvectors(a, b, ev, z, .false., success)             
! now b actually contains inverse of the Cholesky 
! decomposition for the next time with the same b,
! call with .true.
call e%generalized_eigenvectors(a, b, ev, z, .true., success)             
\end{lstlisting}
\end{figure}

\subsection{Generalized EVP problem}
\label{sec:generalized}
The new API now features a dedicated function for solving a generalized eigenvector problem. 
Since in applications 
it is often the case that there is a sequence of generalized EVP 
calculations with different matrices $A$ but the same matrix $B$, we designed the API to take  advantage of this. 
The function \emph{generalized\_eigenvectors} has the parameter \emph{is\_already\_decomposed}. If false, first a Cholesky
decomposition (\ref{eq:cholesky}) of matrix $B$ is computed and the result is then inverted. Both operations are efficiently
implemented inside the ELPA library and are performed in-place, so the matrix $B$ is overwritten on the output of 
\emph{generalized\_eigenvectors} as a side-effect of the function call. For the subsequent calls to the generalized EVP 
solver with the same matrix $B$, this modified value is provided together with setting the parameter \emph{is\_already\_decomposed}
to true, as it is shown in Code~\ref{code:generalized}.

In both cases, an efficient implementation of triangular matrix multiplication based on modification of the Cannon algorithm,
described in \cite{cannon_alg}, is used for the transformations (\ref{eq:trans_gen_forw}) and (\ref{eq:trans_gen_back}).
The authors of \cite{cannon_alg} show that their algorithm has very good scaling
properties. 
Furthermore, even in the case, when only one generalized EVP with a given matrix $B$ is computed and 
thus the overhead of explicitly inverting its Cholesky decomposition is the largest, the total cost of the computation is
almost always smaller than in an alternative approach without its explicit construction. For the algorithm details, performance 
comparisons and further references see  \cite{cannon_alg}.


\subsection{Autotuning}
\label{sec:autotuning}

\begin{figure}
 \begin{lstlisting}[label={code:autotuning}, caption={
Example use of autotuning. In a real-world application, the artificial while cycle can be 
replaced by the existing logic.
}]
use elpa
class(elpa_t),          pointer :: e
class(elpa_autotune_t), pointer :: tune_state
e => elpa_allocate()
! set all the required fields, omitting others
call e%set("na", na, error)
! alternatively exclude some parameters from autotuning by setting them
call e%set("gpu", 0)
!set up the ELPA object and create the autotuning object
success = e%setup()
tune_state => e%autotune_setup(level, domain, error)
iter=0
! instead of this, call e.g. inside the SCF cycle
do while (e%autotune_step(tune_state))
  iter=iter+1
  call e%eigenvectors(a, ev, z, error)
  ! do whatever needed with the result
  if (iter > MAX_ITER) then  ! optionally...
    ! the status of the autotuning can be saved
    call e%autotune_save_state(tune_state, "autotune_checkpoint.txt")
    ! the autotuning then can be stopped and resumed
    exit
  endif
end do
! set and print the autotuned-settings
call e%autotune_set_best(tune_state)
! the current values of the parameters can be saved 
call e%save_all_parameters("autotuned_params.txt")
! de-allocate autotune object
call elpa_autotune_deallocate(tune_state)
\end{lstlisting}
\end{figure}

As we have described above, one of the advantages of the new API is the possibility to add parameters which can influence
the performance for different setups on different hardware. However, a new problem arises from the fact that choosing
the best (w.r.t. performance) values for these parameters either requires expert knowledge of the user or the tedious approach
to test different possibilities. Although in the ELPA library  there are reasonable default values for all parameters (chosen 
such that the performance should be optimal in most cases), it still very much depends on the particular hardware
and the setting of the problem to solve (size of the matrix, Scalapack block-size, number of eigenvectors 
sought, process distribution, etc.), whether the default is delivering the best performance.

For all users that do not want to find the best setting by themselves, we developed an autotuning capability of
the ELPA library, which should allow the user to find the best settings of parameters in a (semi) automated way.
Autotuning works by solving repeatedly the same (or a very similar) problem over and over again, and testing at each step
one of the possible combinations of all parameters. At the end the setting with the best run-time is reported to the user
and can than be used for the subsequent calculations. This situation arises quite naturally in practical applications of
electronic structure theory, where quite often the so-called self-consistent field (SCF) problem is solved iteratively (see Section~\ref{sec:applications}).

In order to make the autotuning functionality as flexible as possible for the user, the ELPA library defines certain sets of 
tunable-parameters. With the set \emph{FAST} the autotuning finishes much faster than with the set 
\emph{MEDIUM}, but of course some possible combinations are not considered in the former set.
The user is offered even more flexibility by excluding some settings from the search tree by manually setting parameters to a
fixed value.
An example of setting up the autotuning is given in the Code~\ref{code:autotuning}, where the tuning of the GPU usage is manually excluded.
If there are not enough iterations within the SCF loop to finish the autotuning process, a snapshot of its current state
can be saved and the process can be resumed later. When the autotuning process is finished and the values of the parameters 
of the ELPA object are satisfactory, they can be also saved to a file for a future use (to have the optimal parameters
without the need of running the autotuning again). Both mentioned possibilities are outlined in Code~\ref{code:autotuning}.

\begin{table*}
\begin{center}
\caption{\label{tab:gpu_autotuning} Performance of individual stages of the ELPA 1 algorithm. 
For each phase (and the total time), the faster variant is shown in bold.}
\begin{tabular}{c|c c|c c|c c|c c c}
\hline
{}& \multicolumn{2}{c|}{TRIDI}  & \multicolumn{2}{c|}{SOLVE} & \multicolumn{2}{c|}{BACK}  & \multicolumn{3}{c}{TOTAL}  \\ 
size  &CPU & GPU  &CPU & GPU  &CPU & GPU     & CPU &GPU & TUNED\\
\hline
1024 & \textbf{0.04} & 0.44 & \textbf{0.01} & 0.06 & \textbf{0.02} & 0.02 & \textbf{0.07} & 0.56 & 0.07 \\  
2048 & \textbf{0.15} & 0.85 & \textbf{0.06} & 0.1  & 0.07 & \textbf{0.05} & 0.28 & 1.0 & \textbf{0.26} \\ 
4096 & \textbf{0.79} & 2.0  & 0.3  & \textbf{0.22} & 0.42 & \textbf{0.13} & 1.5  & 2.4 & \textbf{1.1} \\ 
8192 & 9.5  & \textbf{4.5}  & 1.9  & \textbf{0.74} & 2.9  & \textbf{0.53} & 14.3 & \textbf{5.8} & 5.8  \\ 
\hline
\end{tabular}
\end{center}
\end{table*}
We illustrate the possibilities of autotuning on a particular example.
In Section~\ref{ssec:gpu} we have described the GPU implementation of ELPA and showed that the CPU version 
of ELPA is faster for small matrices and the GPU version is faster for large ones, but the turning point 
is different for individual parts of the algorithm.
%
Since the user can have a full control over which routine runs on CPU and which on GPU, either
directly or using autotuning, 
further improvements can be achieved, as it can be seen from Table~\ref{tab:gpu_autotuning}.
For example, 
for the matrix size 4096, the tuned version runs only 1.1 seconds, which is 1.4x faster than the CPU-only and 2.2x 
faster than the GPU-only version.
%
%

\section{Applications in quantum mechanics}
\label{sec:applications}
The solution of the electronic structure problem is at the basis of any
computation in theoretical chemistry, molecular physics, and computational
materials science. 
The mass of the electron as well as its fermionic nature
require the electronic structure problem to be treated quantum mechanically.

\subsection{Introduction to electronic structure computations}
\label{Intro:QM}
The most common approach introduces a
mean-field potential effectively experienced by each electron moving in the 
field of the remaining $N-1$ electrons and the nuclei. 
This ansatz yields a set of coupled integro-differential equations which need to
be solved iteratively in a process known as the SCF
method until a stationary point is reached, where the mean-field potential
reproduces itself.
This procedure is common to the Hartree-Fock (HF) molecular orbital method, which
forms the basis of all wavefunction-based electronic structure methods, as
well as the Kohn-Sham (KS) orbital method in density functional theory (DFT).

In practice, the integro-differential equations are algebraized by
introducing an appropriate basis in the Hilbert space of admissible single
electron wavefunctions, also called orbitals, from which the many-electron
wavefunctions (configurations) are constructed.
Approximate numerical solutions to the electronic structure
problem are then obtained by truncating the basis at finite size $M \geq N$ 
yielding a generalized matrix eigenvalue problem
$A V =  B V \Lambda$. 
The size of the problem depends on the size
of the basis. 
In contrast to the so-called "overlap-matrix" $B$ (usually termed
$S$ in the electronic structure literature), the "Hamiltonian matrix" $A$
(usually termed $H$) depends itself on the lower part (eigenvalue/-vector pairs) 
of the eigensystem.
A single step consisting of computing the matrices 
$A$ and $B$ 
and solving the eigenproblem is referred to as
an "SCF step", while the fixed point iteration involving a series of such
steps is known as an "SCF cycle". 
The final result of an SCF cycle is the lower part 
of the eigensystem, from which the total energy $E(\vec{R})$ at the given
nuclear configuration $\vec{R}$ can be computed.

$E(\vec{R})$ is known as the potential energy surface (PES) for the 
dynamics of the nuclei and 
determines the chemistry and physics of molecules and materials at
atomistic resolution.
Methods exploring the PES (Minimization, Saddle Point Search, \textit{ab initio}
Molecular Dynamics) thus require an SCF cycle at each geometry $\vec{R}$. 
For an atom-centered basis, the matrix
$B$ will only change for a modification in $\vec{R}$
, while it remains constant for
the whole SCF cycle, during which only the matrix $A$ is updated in each
SCF step. 


\begin{figure}
\centering
\includegraphics[width=0.5\textwidth]{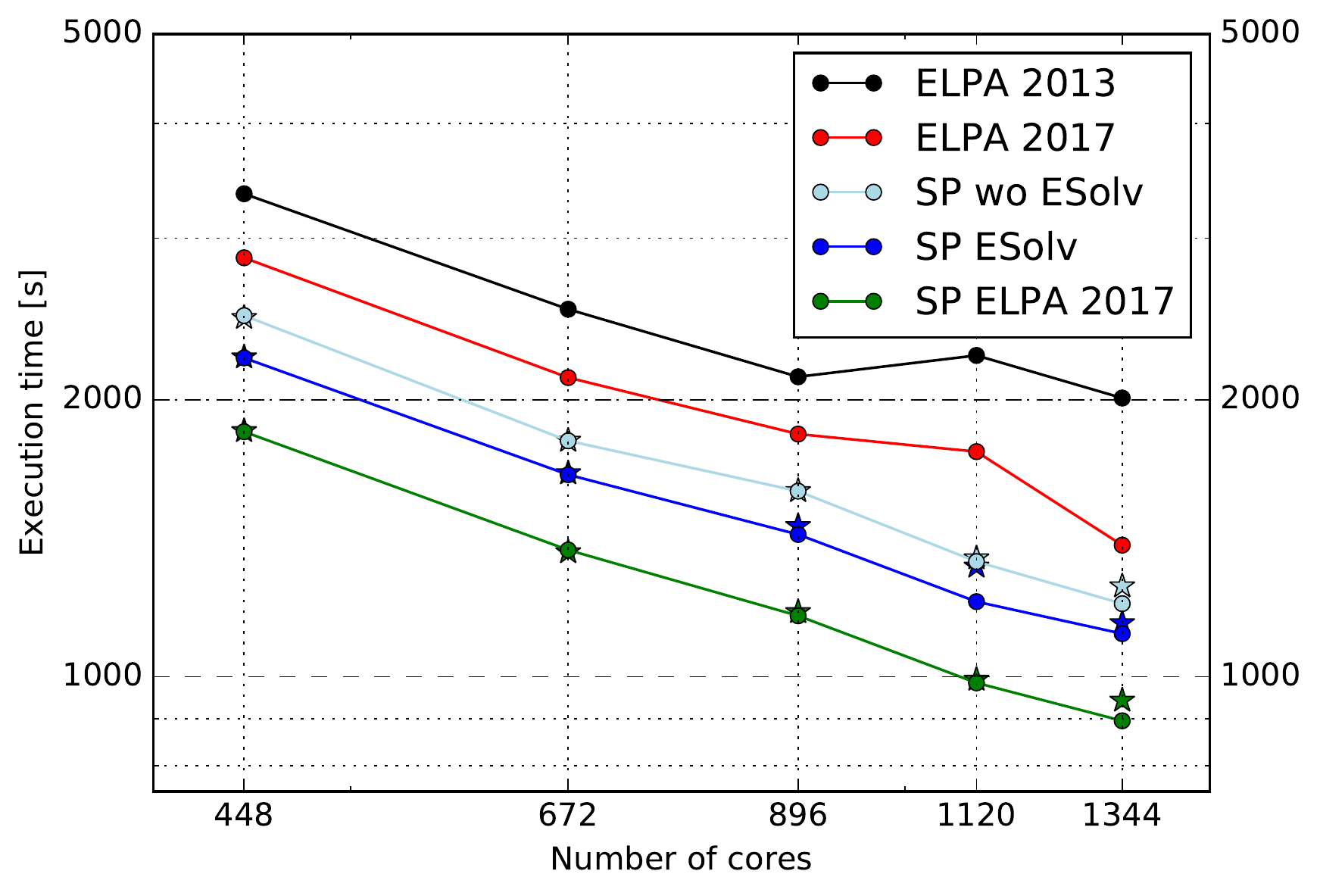}
\caption{
Scaling of computational time for solving the Kohn-Sham eigenvalue problem for a IrO$_2$ nano cluster (87009 basis functions). 
	The FHI-aims internal ELPA 2013 version (black), which was the standard before the beginning of the ELPA-AEO project, is compared to ELPA-AEO 2017 (red). The difference reflects the performance improvements of the ELPA library \emph{without} the AVX-512 improvements discussed in Section~\ref{sec:architecture_optimization}.
SP is applied either in all steps (dark green), only in step (iv) (blue, SP Esolv), or in steps (i)-(iii) (light blue, wo ESolv). 
The conversion of the matrices from DP to SP and vice versa is conducted with method (a) (element by element, o symbols). 
Conversion of the matrix type by method (b) (as block) is depicted by stars.
}
\label{fig:KS_IrO2}
\end{figure}

\begin{figure}
\centering
\includegraphics[width=0.5\textwidth]{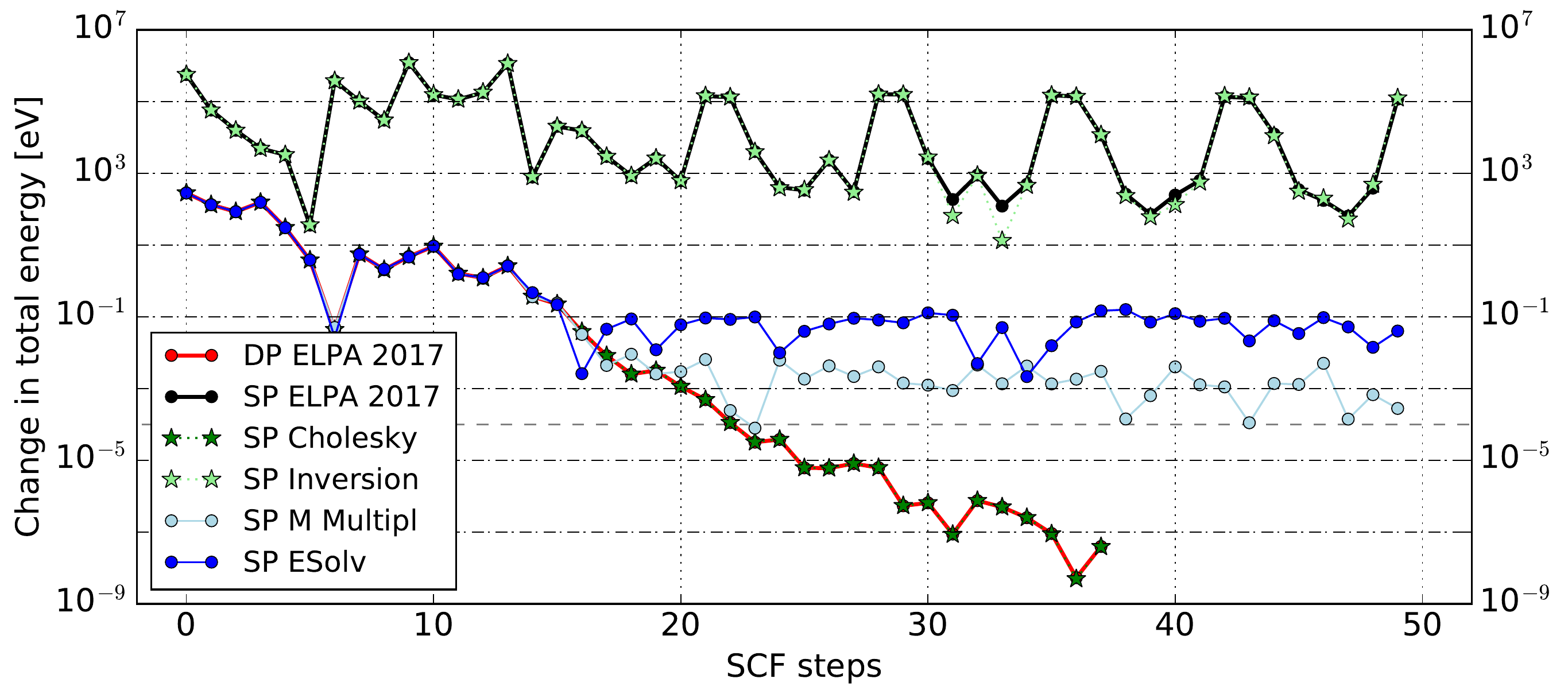}
\caption{
Impact of the precision on the convergence of the total energy for a TiO$_2$ cluster (4784 basis functions). 
DP (red) and SP (black) in all steps are depicted by a full line. 
SP in step (i) (dark green, Cholesky decomposition) or in step (ii) (light green, inversion of overlap matrix) are shown by stars. 
SP in step (iii) (matrix multiplication) or step (iv) (solution of tridiagonal eigenproblem) are shown in light blue and blue, respectively. 
}
\label{fig:conv_TiO2_totE}
\end{figure}
\subsection{Mixing single and double precision}
\label{ssec:mixed_precision}

A typical computational study requires thousands if not millions of SCF cycles~(about 10-100 SCF steps per cycle) at varying geometries $\vec{R}$ to be performed in a single simulation.
This makes it worthwhile to assess strategies to reduce the computational effort. 
Only the final results of the converged SCF cycle are of physical relevance at all. 
Hence, the SCF procedure can be accelerated by using single precision~(SP) routines instead of double precision~(DP) ones in the appropriate eigensolver steps, as long as the final converged result is not altered up to the precision mandated by the problem at hand. 
The eigensolver steps discussed in this section, are the Cholesky decomposition (i), c.f. Eq.(\ref{eq:cholesky}), the inversion of the overlap matrix (ii), and the matrix multiplication (iii) in Eq.~(\ref{eq:trans_gen_forw}), as well as the solution of the eigenproblem via tridiagonalization (iv), c.f. Eqs.~(\ref{eq:tridiagonalization}) and (\ref{eq:Esolv}).

To showcase the importance of the readily available SP routines in ELPA-AEO, we have performed DFT calculations with the all-electron, numeric atomic orbitals based code FHI-aims\cite{Blum:2009fe}, which supports both ELPA and ELPA-AEO through the ELSI package.~\cite{Yu:2018ih}. 
Benchmark calculations for an iridium oxide nano-cluster (IrO$_2$, 1857 atoms) 
quantify the speed-up achieved by different precisions by evaluating two SCF cycles and the atomic forces with 87009 
basis functions.
The calculations were conducted on Intel Xeon processors E5-2697 (28 cores @ 2.6 GHz in 2 CPUs per node). 
Replacing the FHI-aims internal ELPA 2013 (standard before the ELPA-AEO project) by ELPA-AEO 2017 achieves a speed-up of 1.2 for the solution of the Kohn-Sham eigenvalue problem. Please note, that this speed-up is achieved without the AVX-512 optimizations as discussed in Section~\ref{sec:architecture_optimization}. With AVX-512 optimizations enabled in ELPA-AEO 2017 (on the proper hardware) another factor of 1.5 to 2 could be achieved.
Figure \ref{fig:KS_IrO2} shows that SP in all four steps achieves an additional speed-up of 1.5 in Kohn-Sham computational time 
in comparison to DP calculations with ELPA-AEO 2017. 
The speed-up due to SP in the tridiagonal eigensolver (iv) or due to SP in all steps BUT 4 (SP in steps (i)-(iii)) amounts to 1.3 
and 1.1, 
respectively. 
Hence, the tridiagonal eigensolver provides the largest contribution to the speed-up of SP versus DP. 
Copying the matrices from DP to SP and vice versa can be accomplished by two different methods: Either the matrix elements are copied individually (a) or the entire matrix is copied as block (b). 
Method (a) usually requires not as much computational time as method (b) but the gain is marginal. 

In Figure \ref{fig:conv_TiO2_totE}, the convergence of the total energy in dependency on the precision of each step is studied for a TiO$_2$ cluster (4785 basis functions). 
Since FHI-aims expands the electron wavefunctions in an atom-centred basis, the overlap matrix $B$ is only constructed once per SCF cycle and the Cholesky decomposition (i) is only conducted in the first SCF iteration step of each SCF cycle.
The precision of the Cholesky decomposition (i) in the first SCF iteration does not influence the convergence. 
SP in step (iii) or (iv) reduces the convergence of energy and electron density after iteration step 20. 
Reduction of precision for the overlap matrix inversion (ii) destroys the convergence entirely.
Nevertheless, the gain in computational efficiency by SP in step (iii) and (iv) can be exploited during the 20 SCF steps, after which the precision is switched to DP in order to achieve final convergence. 


\subsection{Performance benefits by autotuning}
\label{ssec:autotuning_fhi}
As discussed in Sec.~\ref{sec:autotuning}, using the optimal ELPA settings~(kernel, utilization of GPUs, etc.) can lead to 
additional computational savings in practical calculations. 
In particular, this applies to electronic structure calculations, which involve solving many similar eigenvalue 
problems~(SCF steps) 
in one SCF cycle~(see Sec.~\ref{Intro:QM}). However, identifying these optimal settings, which depend upon both 
the inspected physical 
problem and the used architecture, is a tedious job if performed manually. To take this burden from the user, 
ELPA's new autotuning 
feature allows to determine these settings in an automated fashion. To showcase the impact of this feature 
in practical applications, we have performed DFT benchmark calculations using the FHI-aims~\cite{Blum:2009fe,Yu:2018ih} 
code for two extended periodic systems,~i.e.,~an A-DNA double helix~\cite{LinLin:2014DNA} 
(7150 atoms in the unit cell, 77220 basis functions) and a Graphene sheet (5000 atoms in the unit cell, 70000 
basis functions) using the PBE exchange-correlation functional~\cite{Perdew:PBE}. Two different autotuning presets were
used: Autotuning FAST, in which only the choice of kernel is optimized, and autotuning MEDIUM, in which 
the numerical blocking parameters of the back-transformation are additionally optimized. In both cases, the ELPA 1 solver
was explicitly excluded from the autotuning procedure, since ELPA 2 is superior for these particular problems and architecture.
Also, GPU accelerators and a (hybrid) OpenMP parallelization were not utilised in these calculations that were performed on 32 Intel 
Skylake nodes with 2 CPUs per node each~(20 cores/CPU @ 2.4 GHz).

\begin{figure}
\begin{center}
\centering
\includegraphics[width=0.4\textwidth]{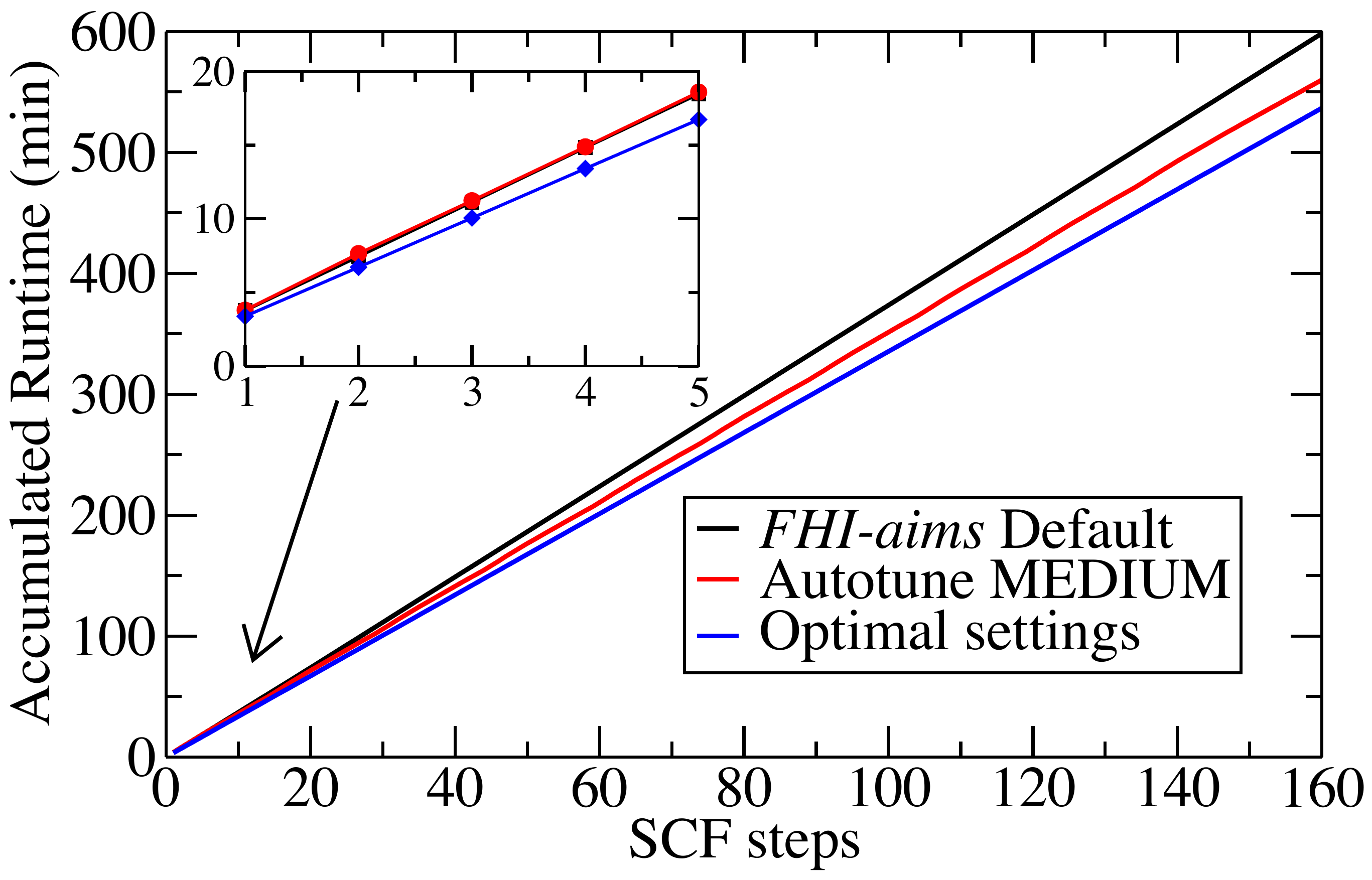}
 \caption{\label{fig:DNAtest} Accumulated runtime (in minutes) 
for A-DNA~(see text) in three different scenarios: Using FHI-aims' default settings, using autotuning level MEDIUM, 
and using optimal settings,
as if these were known from the start. 
There is a clear benefit of autotuned version over the FHI-aims default one. Moreover, if more SCF iterations were done, 
the \emph{relative} difference between the autotuned and optimal version would be further reduced.
}
\end{center}
\end{figure}

To qualitatively discuss the impact of autotuning, we compare the accumulated run-times for A-DNA 
in three different scenarios in Fig.~\ref{fig:DNAtest}: Using the default fallback settings~(Generic Kernel)
chosen by FHI-aims if no kernel is specified by the user, using
autotuning level MEDIUM, and using the optimal settings identified by autotuning level MEDIUM, 
as if they were known from the beginning. In the first steps, we see that 
autotuning can be even slower than the generic kernel, given that the
\textit{settings, which are also non-optimal in these first steps} 
are explored. For the exact same reason, the calculation with autotuning level MEDIUM is also
slower than the one using optimal settings. After 150 SCF steps, the optimal settings are
identified in the autotuning level MEDIUM run and the accumulated run-times now show the exact same
slope as the run using optimal settings; the offset between the two curves quantifies the 
cost of the autotuning procedure. Compared to the calculation with default settings, the
computational savings in total runtime are in the order of 5\%. The behaviour described 
above is also observed for the graphene system. In that case, the computational savings 
using autotuning level MEDIUM are in the order of 10\% after 160 SCF steps, as summarized 
in Tab.~\ref{tab:cobratuning}. A qualitative similar behaviour is also observed when only the choice 
of the kernel is optimized~(autotuning level FAST). In this case, only 15 iterations are needed 
to find the optimal kernel~(AVX512).

\begin{table}
\begin{center}
\caption{\label{tab:cobratuning} Average runtime per SCF step 
for the A-DNA (77220 basis functions) and Graphene (70000 basis functions) systems for different kernels and autotuning methods.
Autotuning level FAST requires 15 SCF steps~(20 steps performed in total) to identify the optimal kernel~(AVX-512), 
while autotuning level MEDIUM requires 150 SCF steps~(160 steps performed in total). 
As a reference, timings for using FHI-aims' default settings~(Generic kernel) and for using the optimal settings from the start,~i.e.,~the 
ones identified by autotuning MEDIUM, are given.}
\begin{tabular}{c|c|c|c|c}
\hline
 System  &Generic & Optimal  & FAST & MEDIUM \\
\hline
A-DNA    & 221.3 s & 200.5 s & 209.2 s & 209.6 s  \\  
Graphene & 160.8 s & 137.0 s & 143.5 s &  144.4 s \\
\hline
\end{tabular}
\end{center}
\end{table}


Note that the number of SCF steps required to identify the optimal settings is rather high~(150 for autotune level MEDIUM and 15 for FAST) and thus larger than the typical number of SCF steps~(10-30 in non-problematic systems) needed to achieve convergence in a single SCF cycle. In this light, the computational gain achieved by the autotuning feature for a single SCF cycle might seem negligible. However, almost all electronic structure calculations do not only require one SCF cycle, but many. In practice, hundreds~(structure optimisations) or even several millions~(\textit{ab initio} Molecular Dynamics simulations) of SCF cycles are performed in such applications, whereby the inspected geometry and thus the structure of the eigenvalue problem is only slightly altered between the different SCF cycles. Due to this fact, the optimal settings identified by the autotuning are transferable across SCF cycles, as demonstrated below. 
Accordingly, the autotuning functionality leads to considerable performance benefits in these applications, since the optimal settings --once 
they are identified after the first SCF steps-- can be applied to all further SCF steps and cycles.

\begin{figure}
\begin{center}
\centering
\includegraphics[width=0.4\textwidth]{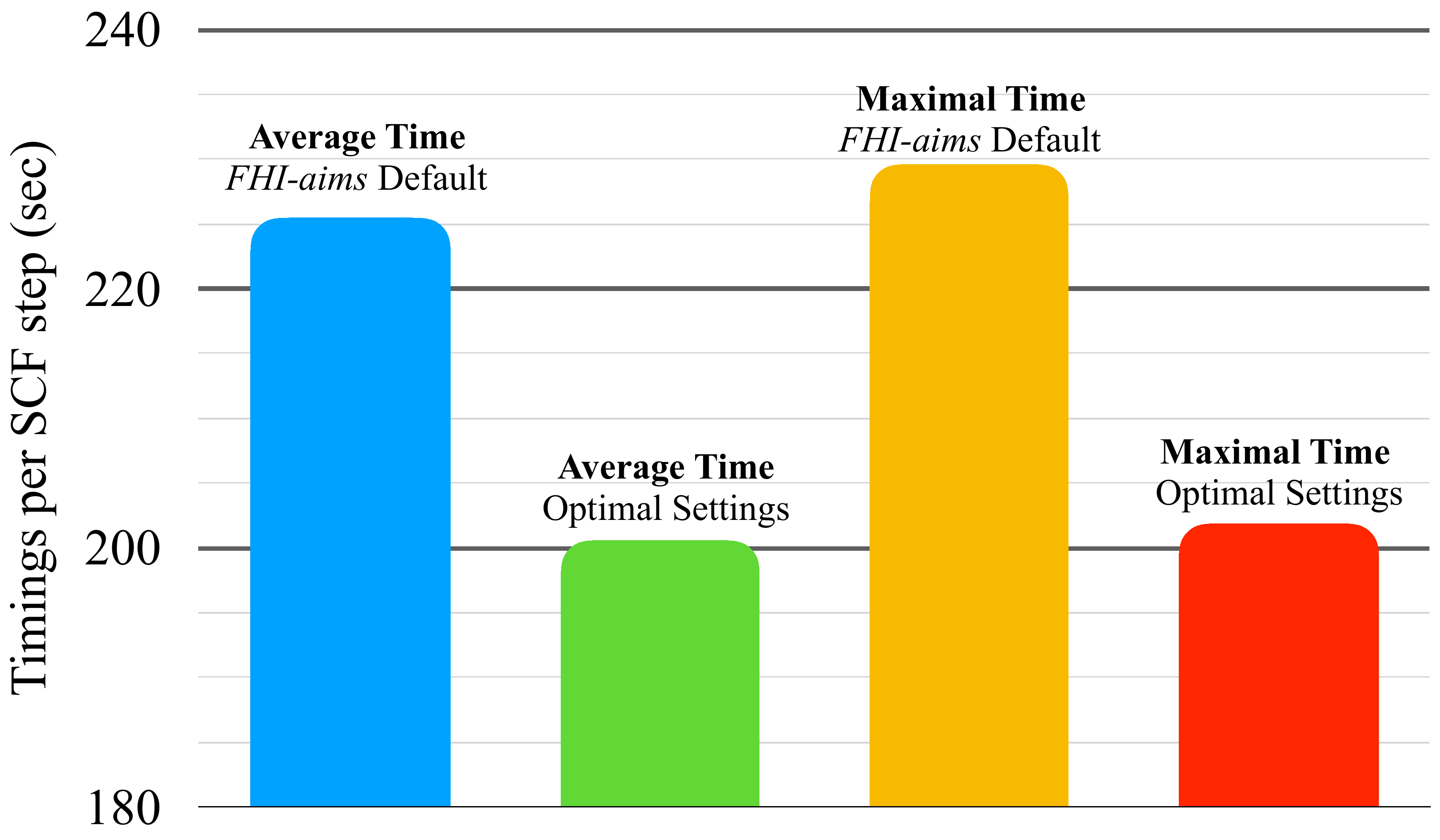}
 \caption{\label{fig:stats} 
Average and maximum runtime per SCF steps observed for ten different geometries of A-DNA~(see text), as they would be
observed in structure relaxations or \emph{ab initio} Molecular Dynamics simulations. Here, we compare timings for
the default and the optimal settings.}
\end{center}
\end{figure}

To showcase these savings, we have inspected ten additional, slightly different geometries of A-DNA, which were generated by randomly displacing
atoms by fractions of~$\mbox{\AA}$~\cite{ASE}. Such geometries are representative for the typical variances in geometry
that are observed in structure relaxations and \emph{ab initio} Molecular Dynamics simulations. We then performed DFT
calculations~(one SCF cycle with approximately 20 SCF steps) for these geometries using (a)~FHI-aims' default settings~(Generic 
Kernel) and (b)~the optimal settings identified
by autotuning level MEDIUM for the equilibrium geometry discussed in Fig.~\ref{fig:DNAtest}. Note that this equilibrium geometry 
is not part of the ten geometries inspected here. Fig.~\ref{fig:stats} demonstrates that the optimal settings found by
autotuning are indeed transferable across SCF cycles and that the associated computational savings are retained along multiple
SCF cycles. This has been additionally verified by running calculations with enabled autotuning for all ten geometries. 
Compared to the default settings, we indeed observe average savings of approximately~\textbf{10}\% for the average 
and maximum run-times per SCF step. This shows that in extended calculations featuring hundreds of SCF cycles and thousands of SCF steps
the relative small overhead required by autotuning~(see Fig.~\ref{fig:DNAtest}) for identifying the optimal settings in the 
first few SCF steps is indeed negligible. Once the optimal settings are identified, they can be applied to all following SCF 
steps, thus leading to significant performance gains in these applications. Similarly, this shows that also the autotuning
procedure itself can be performed over multiple SCF cycles. This allows an even more extensive search for the optimal settings,
e.g.,~by additionally including  GPU and/or OpenMP parallelization, in practical calculations. In these applications, this can 
be straightforwardly realized by exploiting ELPA's capability to save and load (intermediate and final) 
snapshots of the autotuning status, as discussed for the code example~\ref{code:autotuning} in Sec.~\ref{sec:autotuning}.

 
\section{Conclusions}
\label{sec:conclusions}
We have presented the improvements of the ELPA library made during the ELPA-AEO project and showed ways, how the applications
using ELPA can significantly reduce their time-to-solution. On the one hand we demonstrated significant performance improvements by 
adapting the ELPA library to modern architectures. These hardware-related optimizations for Intel Xeon Skylake processors lead to
a speedup of roughly 1.5 to 2 compared to the Intel Ivy Bridge machine. Furthermore,
on an IBM Power 8 system, equipped with 4 NVIDIA Tesla P100 GPUs a speedup 
(depending on the problem size) up to a factor of roughly 12 compared to a CPU only version could be demonstrated. 

While also in the future the ELPA library will be
continuously optimized for the newest hardware, 
it is becoming increasingly complex to optimize the ELPA library
for different kind of HPC systems: the parameter space of algorithmic choices and performance relevant settings (like block-sizes)
is growing quite rapidly. Although during the development of ELPA we try to define reasonable default values for all performance relevant settings,
we realized that these might not always be the best choice, but in the same time the users of the ELPA library might not want to dive into the
tedious work to find the best settings for their combinations of problems and available hardware. Thus, in this paper we 
presented a new autotuning functionality of the ELPA-library which allows finding of the best settings in an easy and automated way.
By showing examples from scientifically interesting applications, we demonstrated that the autotuning does indeed find the best settings. 
The overhead introduced by the autotuning process in the first hundred diagonalizations is negligible in practical calculations, which typically 
feature thousands of matrix diagonalizations that can all benefit from the best settings identified by the autotuning.
There is a possibility to split the autotuning process into several phases by saving the status of the autotuning
and also a possibility to save the optimal found parameter combination for a future use.

We also discussed an example where the autotuning algorithm is able to determine which parts of the calculations should be offloaded
to GPUs. We discussed that this feature is especially useful for matrix sizes of roughly 4000, where it strongly depends on the used hardware, whether
an acceleration with GPUs is possible or not.
The autotuning mechanism could only be implemented by introducing a new, general API for the ELPA library. In addition to
the autotuning we showed that this API has several other important advantages: for the users it is easily possible to define at the same time 
multiple instances of ELPA (and even autotune them independently) if different problems are to be solved within one application. As an example we
showed a real-world example where a mixed-precision approach was taken and parts of the steps to solve a generalized eigenvalue problem were done 
in single instead of double precision. 


\section*{Acknowledgments}
Part of this work is co-funded by BMBF grant 01IH15001 of the German Government.
\FloatBarrier

\end{document}